\documentclass{ws-procs975x65}

\begin{document}

\title{The evolution of the Universe}
\author{Juan Garcia-Bellido}
\address{Theory Division CERN, CH-1211 Gen\`eve, Switzerland\\
%Departamento de F\'{\i}sica Te\'orica, Universidad Aut\'onoma
%de Madrid, \\ Cantoblanco 28049 Madrid, Spain\\
E-mail: bellido@mail.cern.ch}


\maketitle

\abstracts{With the recent measurements of temperature and polarization
anisotropies in the microwave background by WMAP, we have entered a new
era of precision cosmology, with the cosmological parameters of a
Standard Cosmological Model determined to 1\%. This Standard Model is
based on the Big Bang theory and the inflationary paradigm, a period of
exponential expansion in the early universe responsible for the
large-scale homogeneity and spatial flatness of our observable patch of
the Universe. The spectrum of metric perturbations, seen in the
microwave background as temperature anisotropies, were produced during
inflation from quantum fluctuations that were stretched to cosmological
size by the expansion, and later gave rise, via gravitational collapse,
to the observed large-scale structure of clusters and superclusters of
galaxies. Furthermore, the same theory predicts that all the matter and
radiation in the universe today originated at the end of inflation from
an explosive production of particles that could also have been the
origin of the present baryon asymmetry, before the universe reached
thermal equilibrium at a very large temperature. From there on, the
universe cooled down as it expanded, in the way described by the
standard hot Big Bang model.}

\section{Introduction}

Our present understanding of the universe is based upon the successful
hot Big Bang theory, which explains its evolution from the first
fraction of a second to our present age, around 13 billion years
later. This theory rests upon four strong pillars, a theoretical
framework based on general relativity, as put forward by Albert
Einstein and Alexander A. Friedmann in the 1920s, and three strong
observational facts. First, the expansion of the universe, discovered
by Edwin P. Hubble in the 1930s, as a recession of galaxies at a speed
proportional to their distance from us. Second, the relative abundance
of light elements, explained by George Gamow in the 1940s, mainly that
of helium, deuterium and lithium, which were cooked from the nuclear
reactions that took place at around a second to a few minutes after
the Big Bang, when the universe was a hundred times hotter than the
core of the sun. Third, the cosmic microwave background (CMB), the
afterglow of the Big Bang, discovered in 1965 by Arno A. Penzias and
Robert W. Wilson as a very isotropic blackbody radiation at a
temperature of about 3 degrees Kelvin, emitted when the universe was
cold enough to form neutral atoms, and photons decoupled from matter,
380\,000 years after the Big Bang. Today, these observations are
confirmed to within a few percent accuracy, and have helped establish
the hot Big Bang as the preferred model of the universe.

The Big Bang theory could not explain, however, the origin of matter
and structure in the universe; that is, the origin of the
matter--antimatter asymmetry, without which the universe today would
be filled by a uniform radiation continuosly expanding and cooling,
with no traces of matter, and thus without the possibility to form
gravitationally bound systems like galaxies, stars and planets that
could sustain life. Moreover, the standard Big Bang theory assumes,
but cannot explain, the origin of the extraordinary smoothness and
flatness of the universe on the very large scales seen by the
microwave background probes and the largest galaxy catalogs. It can
neither explain the origin of the primordial density perturbations
that gave rise, via gravitational collapse, to cosmic structures like
galaxies, clusters and superclusters; nor the nature of the dark
matter and dark energy that we believe permeates the universe; nor the
origin of the Big Bang itself.

\begin{figure}[htbp]
\begin{center}
%\vspace*{-.5cm}
\hspace*{-.1cm}
\leavevmode\epsfysize=7cm \epsfbox{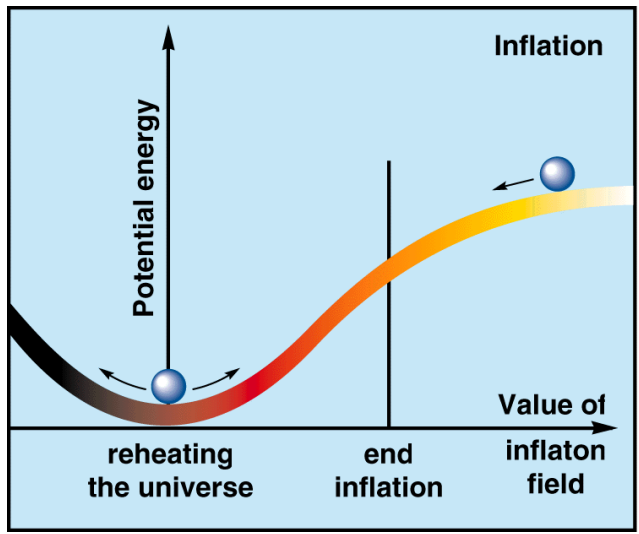}
%\vspace*{1cm}
\caption[fig2]{The inflaton field can be represented as a ball rolling
down a hill.  During inflation, the energy density is approximately
constant, driving the tremendous expansion of the universe. When the
ball starts to oscillate around the bottom of the hill, inflation ends
and the inflaton energy decays into particles. In certain cases, the
coherent oscillations of the inflaton could generate a resonant
production of particles which soon thermalize, reheating the universe.}
\label{fig2}
\end{center}
\end{figure}

In the 1980s, a new paradigm, deeply rooted in fundamental physics,
was put forward by Alan H. Guth, Andrei D. Linde and others, to
address these fundamental questions. According to the inflationary
paradigm, the early universe went through a period of exponential
expansion, driven by the approximately constant energy density of a
scalar field called the inflaton. In modern physics, elementary
particles are represented by quantum fields, i.e. a function of space
and time whose quantum oscillations can be interpreted as
particles. For instance, the photon is the particle associated with
the electromagnetic field. In our case, the inflaton field has,
associated with it, a large potential energy density, which drives the
exponential expansion during inflation, see Fig.~\ref{fig2}. We
know from general relativity that the density of matter determines the
expansion of the universe, but a constant energy density acts in a
very peculiar way: as a repulsive force that makes any two points in
space separate at exponentially large speeds. (This does not violate
the laws of causality because there is no information carried along in
the expansion, it is simply the stretching of space-time.)

\begin{figure}[htbp]
\begin{center}
\vspace*{-4cm}
\hspace*{-1.3cm}
\leavevmode\epsfysize=14.1cm \epsfbox{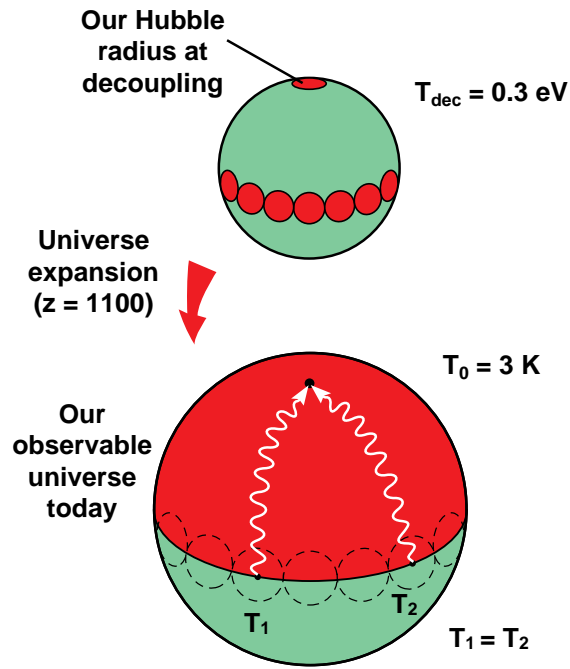}
\vspace*{-1cm}
\caption[fig3]{Perhaps the most acute problem of the Big Bang model is
explaining the extraordinary homogeneity and isotropy of the microwave
background.  Information cannot travel faster than the speed of light,
so the causal region (so-called horizon or Hubble radius) at the time of
photon decoupling could not be larger than 300\,000 light years across,
or about $1^\circ$ projected in the sky today. So why should regions
that are separated by more than $1^\circ$ in the sky have the same
temperature, when the photons that come from those two distant regions
could not have been in causal contact when they were emitted? This
constitutes the so-called horizon problem, which is spectacularly solved
by inflation.}
\label{fig3}
\end{center}
\end{figure}

\begin{figure}[htbp]
\begin{center}
\vspace*{-2cm}
\hspace*{-1cm}
\leavevmode\epsfysize=18cm \epsfbox{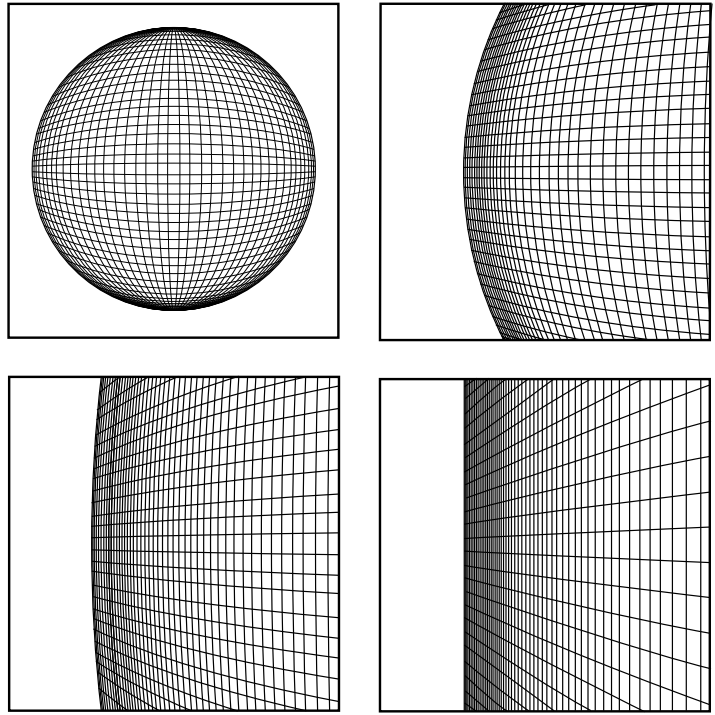}
\vspace*{-6cm}
\caption[fig5]{The exponential expansion during inflation made the
radius of curvature of the universe so large that our observable patch
of the universe today appears essentialy flat, analogous (in three
dimensions) to how the surface of a balloon appears flatter and flatter
as we inflate it to enormous sizes. This is a crucial prediction of
cosmological inflation that will be tested to extraordinary accuracy in
the next few years.}
\label{fig5}
\end{center}
\end{figure}

This superluminal expansion is capable of explaining the large scale
homogeneity of our observable universe and, in particular, why the
microwave background looks so isotropic: regions separated today by
more than $1^\circ$ in the sky were, in fact, in causal contact before
inflation, but were stretched to cosmological distances by the
expansion, see Fig.~\ref{fig3}. Any inhomogeneities present before
the tremendous expansion would be washed out. Moreover, in the usual
Big Bang scenario a flat universe, one in which the gravitational
attraction of matter is exactly balanced by the cosmic expansion, is
unstable under perturbations: a small deviation from flatness is
amplified and soon produces either an empty universe or a collapsed
one. For the universe to be nearly flat today, it must have been
extremely flat at nucleosynthesis for example, deviations not
exceeding more than one part in $10^{15}$. This extreme fine tuning of
initial conditions was also solved by the inflationary paradigm, see
Fig.~\ref{fig5}. Thus inflation is an extremely elegant hypothesis
that explains how a region much, much greater that our own observable
universe could have become smooth and flat without recourse to 
{\em ad hoc} initial conditions.

\section{The origin of structure in the universe}

If cosmological inflation made the universe so extremely flat and
homogeneous, where did the galaxies and clusters of galaxies come from?
One of the most astonishing predictions of inflation, one that was not
even expected, is that quantum fluctuations of the inflaton field are
stretched by the exponential expansion and generate large-scale
perturbations in the metric. Inflaton fluctuations are small wave
packets of energy that, according to general relativity, modify the
space-time fabric, creating a whole spectrum of curvature
perturbations. The use of the word spectrum here is closely related to
the case of light waves propagating in a medium: a spectrum
characterizes the amplitude of each given wavelength. In the case of
inflation, the inflaton fluctuations induce waves in the space-time
metric that can be decomposed into different wavelengths, all with
approximately the same amplitude, that is, corresponding to a
scale-invariant spectrum. These patterns of perturbations in the metric
are like fingerprints that unequivocally characterize a period of
inflation. When matter fell in the troughs of these waves, it created
density perturbations that collapsed gravitationally to form galaxies,
clusters and superclusters of galaxies, with a spectrum that is also
scale invariant. Such a type of spectrum was proposed in the early 1970s
(before inflation) by Edward R. Harrison, and independently by the
Russian cosmologist Yakov B. Zel'dovich, to explain the distribution of
galaxies and clusters of galaxies on very large scales in our observable
universe.

Various telescopes -- like the Hubble Space Telescope, the twin Keck
telescopes in Hawaii and the European Southern Observatory telescopes
in Chile -- are exploring the most distant regions of the universe and
discovering the first galaxies at large distances.  According to the
Big Bang theory, the further the galaxy is, the larger its recession
velocity, and the larger the shift towards the red of the spectrum of
light from that galaxy. Astronomers thus measure distances in units of
red-shift $z$. The furthest galaxies observed so far are at redshifts
of $z\simeq7$, or 13 billion light years from the Earth, whose light
was emitted when the universe had only about 2\% of its present age.
Only a few galaxies are known at those redshifts, but there are at
present various catalogs like the IRAS PSCz and Las Campanas redshift
survey, that study the spatial distribution of hundreds of thousands
of galaxies up to distances of a billion light years, or $z<0.1$, that
recede from us at speeds of tens of thousands of kilometres per
second. These catalogs are telling us about the evolution of clusters
of galaxies in the universe, and already put constraints on the theory
of structure formation based on the gravitational collapse of the
small inhomogeneities produced during inflation. From these
observations one can infer that most galaxies formed at redshifts of
the order of $2 - 4$; clusters of galaxies formed at redshifts of
order 1, and superclusters are forming now. That is, cosmic structure
formed from the bottom up: from galaxies to clusters to superclusters,
and not the other way around.

This fundamental difference is an indication of the type of matter
that gave rise to structure. We know from primordial nucleosynthesis
that all the baryons in the universe cannot account for the observed
amount of matter, so there must be some extra matter (dark since we
don't see it) to account for its gravitational pull. Whether it is
relativistic (hot) or non-relativistic (cold) could be inferred from
observations: relativistic particles tend to diffuse from one
concentration of matter to another, thus transferring energy among
them and preventing the growth of structure on small scales. This is
excluded by observations, so we conclude that most of the matter
responsible for structure formation must be cold. How much there is is
a matter of debate at the moment. Some recent analyses suggest that
there is not enough cold dark matter to reach the critical density
required to make the universe flat. If we want to make sense of the
present observations, we must conclude that some other form of energy
permeates the universe. In order to resolve this issue, even deeper
galaxy redshift catalogs are underway, looking at millions of
galaxies, like the Sloan Digital Sky Survey (SDSS) and the
Anglo-Australian two degree field Galaxy Redshift Survey, which
are at this moment taking data, up to redshifts of $z<3$, or several
hundred billion light years away, over a large region of the sky. 
These important observations will help astronomers determine the
nature of the dark matter and test the validity of the models of
structure formation.

\begin{table}\label{table1}
\caption{{\bf The parameters of the standard cosmological model}. The
standard model of cosmology has about 20 different parameters, needed
to describe the background space-time, the matter content and the
spectrum of metric perturbations. We include here the present range of
the most relevant parameters (with 1$\sigma$ errors), as recently
determined by MAP, and the error with which the Planck satellite will
be able to determine them in the near future.  The rate of expansion
is written in units of $H=100\,h$ km/s/Mpc}
\begin{tabular}{lllll}
\hline
physical quantity & 
\multicolumn{1}{c}{symbol} &
\multicolumn{1}{c}{MAP} &
\multicolumn{1}{c}{Planck} \\
\hline
\hline
total density & 
\multicolumn{1}{c}{$\Omega_0$} &
\multicolumn{1}{c}{$1.02\pm0.02$} &
\multicolumn{1}{c}{0.7\%} \\
\hline
baryonic matter & 
\multicolumn{1}{c}{$\Omega_{\rm B}$} &
\multicolumn{1}{c}{$0.044\pm0.004$} &
\multicolumn{1}{c}{0.6\%} \\
\hline
cosmological constant & 
\multicolumn{1}{c}{$\Omega_\Lambda$} &
\multicolumn{1}{c}{$0.73\pm0.04$} &
\multicolumn{1}{c}{0.5\%} \\
\hline
cold dark matter & 
\multicolumn{1}{c}{$\Omega_{\rm M}$} &
\multicolumn{1}{c}{$0.23\pm0.04$} &
\multicolumn{1}{c}{0.6\%} \\
\hline
hot dark matter & 
\multicolumn{1}{c}{$\Omega_\nu h^2$} &
\multicolumn{1}{c}{$<0.0076$ (95\% c.l.)} &
\multicolumn{1}{c}{1\%} \\
\hline
sum of neutrino masses & 
\multicolumn{1}{c}{$\sum m_\nu$ (eV)} &
\multicolumn{1}{c}{$<0.23$ (95\% c.l.)} &
\multicolumn{1}{c}{1\%} \\
\hline
CMB temperature & 
\multicolumn{1}{c}{$T_0\ (K)$} &
\multicolumn{1}{c}{$2.725\pm0.002$} &
\multicolumn{1}{c}{0.1\%} \\
\hline
baryon to photon ratio & 
\multicolumn{1}{c}{$\eta\times10^{10}$} &
\multicolumn{1}{c}{$6.1\pm0.3$} &
\multicolumn{1}{c}{0.5\%} \\
\hline
baryon to matter ratio & 
\multicolumn{1}{c}{$\Omega_{\rm B}/\Omega_{\rm M}$} &
\multicolumn{1}{c}{$0.17\pm0.01$} &
\multicolumn{1}{c}{1\%} \\
\hline
spatial curvature & 
\multicolumn{1}{c}{$\Omega_K$} &
\multicolumn{1}{c}{$<0.02$ (95\% c.l.)} &
\multicolumn{1}{c}{0.5\%} \\
\hline
rate of expansion & 
\multicolumn{1}{c}{$h$} &
\multicolumn{1}{c}{$0.71\pm0.03$} &
\multicolumn{1}{c}{0.8\%} \\
\hline
age of the universe & 
\multicolumn{1}{c}{$t_0$ (Gyr)} &
\multicolumn{1}{c}{$13.7\pm0.2$} &
\multicolumn{1}{c}{0.1\%} \\
\hline
age at decoupling & 
\multicolumn{1}{c}{$t_{\rm dec}$ (kyr)} &
\multicolumn{1}{c}{$379\pm8$} &
\multicolumn{1}{c}{0.5\%} \\
\hline
age at reionization & 
\multicolumn{1}{c}{$t_{\rm r}$ (Myr)} &
\multicolumn{1}{c}{$180\pm100$} &
\multicolumn{1}{c}{5\%} \\
\hline
spectral amplitude & 
\multicolumn{1}{c}{$A$} &
\multicolumn{1}{c}{$0.833\pm0.085$} &
\multicolumn{1}{c}{0.1\%} \\
\hline
spectral tilt (at $k_0=0.05$ Mpc$^{-1})$& 
\multicolumn{1}{c}{$n_{\rm s}$} &
\multicolumn{1}{c}{$0.93\pm0.03$} &
\multicolumn{1}{c}{0.2\%} \\
\hline
spectral tilt variation& 
\multicolumn{1}{c}{$dn_{\rm s}/d\ln k$} &
\multicolumn{1}{c}{$-0.031\pm0.017$} &
\multicolumn{1}{c}{0.5\%} \\
\hline
tensor-scalar ratio & 
\multicolumn{1}{c}{$r$} &
\multicolumn{1}{c}{$<0.71$ (95\% c.l.)} &
\multicolumn{1}{c}{5\%} \\
\hline
reionization optical depth & 
\multicolumn{1}{c}{$\tau$} &
\multicolumn{1}{c}{$0.17\pm0.04$} &
\multicolumn{1}{c}{5\%} \\
\hline
redshift of matter-energy equality & 
\multicolumn{1}{c}{$z_{\rm eq}$} &
\multicolumn{1}{c}{$3233\pm200$} &
\multicolumn{1}{c}{5\%} \\
\hline
redshift of decoupling & 
\multicolumn{1}{c}{$z_{\rm dec}$} &
\multicolumn{1}{c}{$1089\pm1$} &
\multicolumn{1}{c}{0.1\%} \\
\hline
width of decoupling & 
\multicolumn{1}{c}{$\Delta z_{\rm dec}$} &
\multicolumn{1}{c}{$195\pm2$} &
\multicolumn{1}{c}{1\%} \\
\hline
redshift of reionization & 
\multicolumn{1}{c}{$z_{\rm r}$} &
\multicolumn{1}{c}{$20\pm10$} &
\multicolumn{1}{c}{2\%} \\
\hline

\end{tabular}
\end{table}

\begin{figure}[htbp]
\begin{center}
%\vspace*{-.1cm}
%\hspace*{-1cm}
\leavevmode\epsfysize=6cm \epsfbox{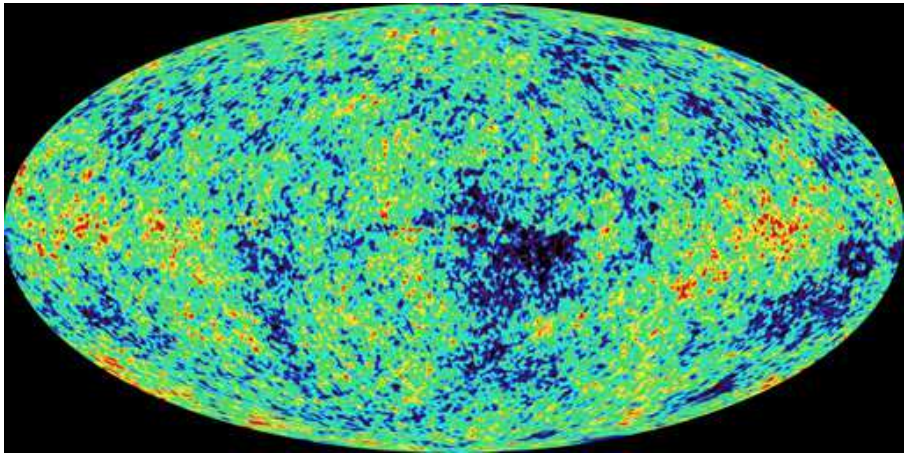}
%\vspace*{-1cm}
\caption[fig6]{The microwave background sky as seen by WMAP, with 10
arc minute resolution. It shows the intrinsic CMB anisotropies,
corresponding to the quadrupole and higher multipoles, at the level of
a few parts in $10^5$. The galaxy is a foreground and has been
subtracted.}
\label{fig6}
\end{center}
\end{figure}

However, if galaxies did indeed form from gravitational collapse of
density perturbations produced during inflation, one should also
expect to see such ripples in the metric as temperature anisotropies
in the cosmic microwave background, that is, minute deviations in the
temperature of the blackbody spectrum when we look at different
directions in the sky. Such anisotropies had been looked for ever
since Penzias and Wilson's discovery of the CMB, but had eluded all
detection, until NASA's Cosmic Background Explorer (COBE) satellite
discovered them in 1992. The reason why they took so long to be
discovered was that they appear as perturbations in temperature of
only one part in 100\,000. There is, in fact, a dipolar anisotropy of
one part in 1000, in the direction of the Virgo cluster, but that is
interpreted consistently as our relative motion with respect to the
microwave background due to the local distribution of mass, which
attracts us gravitationally towards the Virgo cluster. When
subtracted, we are left with a whole spectrum of anisotropies in the
higher multipoles (quadrupole, octopole, etc.), see Fig.~\ref{fig6}. 
Soon after COBE, other groups quickly confirmed the
detection of temperature anisotropies at around 30\,$\mu$K, at higher
multipole numbers or smaller angular scales.

\begin{figure}[htbp]
\begin{center}
\vspace*{-.5cm}
\hspace*{-.5cm}
\leavevmode\epsfysize=6.6cm \epsfbox{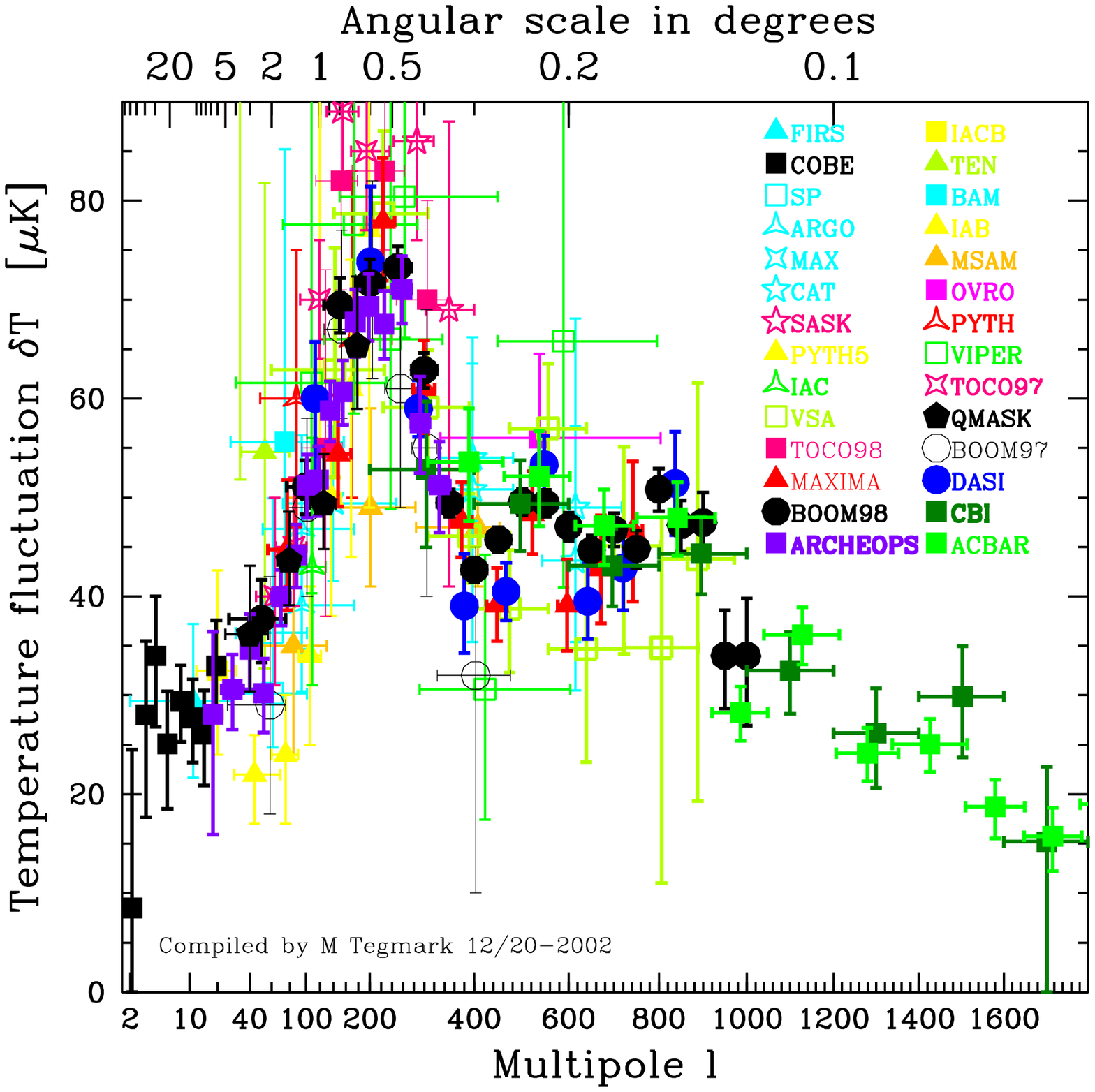}
\hspace*{-.5cm}
\leavevmode\epsfysize=6.6cm \epsfbox{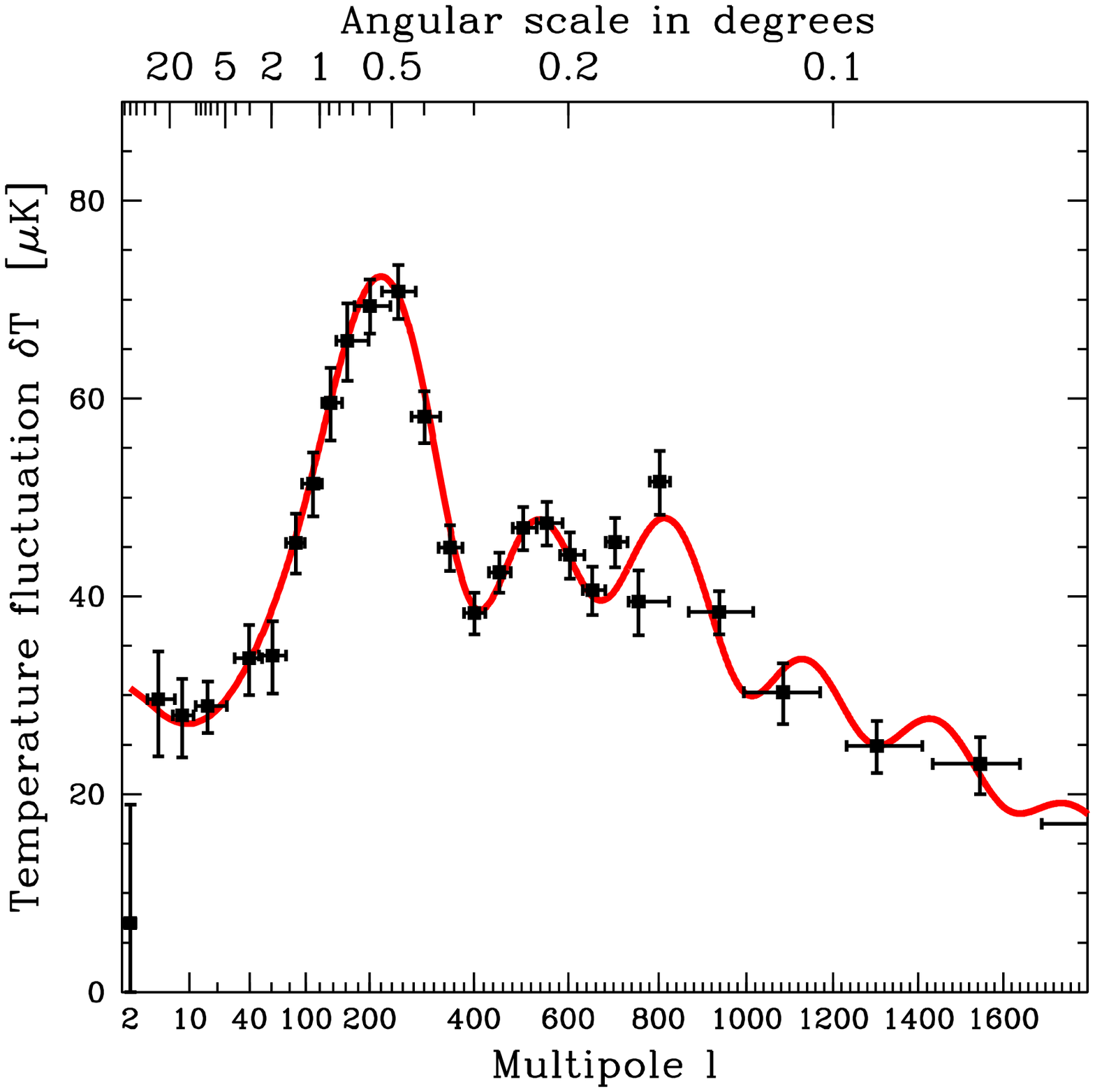}
\vspace*{-.5cm}
\caption[fig7]{There are at present about thirty experiments (in
satellites, from the ground and balloon-borne) looking at the
microwave background temperature anisotropies with angular resolutions
from $7^\circ$ to a few arc minutes in the sky, corresponding to
multipole numbers $l=2 - 3000$. The right panel shows the $l$-binned
spectrum. Present observations suggest the existence of a series of
acoustic peaks in the angular distribution, as predicted by inflation.
The theoretical curve (red thick line) illustrates the concordance
$\Lambda$-CDM model which fits the data.}
\label{fig7}
\end{center}
\end{figure}

There are at this moment dozens of ground and balloon-borne
experiments analysing the anisotropies in the microwave background
with angular resolutions from $7^\circ$ to a few arc minutes in the
sky. The physics of the CMB anisotropies is relatively simple: photons
scatter off charged particles (protons and electrons), and carry
energy, so they feel the gravitational potential associated with the
perturbations imprinted in the metric during inflation. An overdensity
of baryons (protons and neutrons) does not collapse under the effect
of gravity until it enters the causal Hubble radius. The perturbation
continues to grow until radiation pressure opposes gravity and sets up
acoustic oscillations in the plasma, very similar to sound
waves. Since overdensities of the same size will enter the Hubble
radius at the same time, they will oscillate in phase. Moreover, since
photons scatter off these baryons, the acoustic oscillations occur
also in the photon field and induces a pattern of peaks in the
temperature anisotropies in the sky, at different angular scales, see
Fig.~\ref{fig7}. The larger the amount of baryons, the higher the
peaks. The first peak in the photon distribution corresponds to
overdensities that have undergone half an oscillation, that is, a
compression, and appear at a scale associated with the size of the
sonic horizon at last scattering (when the photons decoupled) or about
$1^\circ$ in the sky. Other peaks occur at harmonics of this,
corresponding to smaller angular scales. Since the amplitude and
position of the primary and secondary peaks are directly determined by
the sound speed (and, hence, the equation of state) and by the
geometry and expansion of the universe, they can be used as a powerful
test of the density of baryons and dark matter, and other cosmological
parameters.

By looking at these patterns in the anisotropies of the microwave
background, cosmologists can determine not only the cosmological
parameters but also the primordial spectrum of metric perturbations
produced during inflation. It turns out that the observed temperature
anisotropies are compatible with a scale-invariant spectrum, as
predicted by inflation. This is remarkable, and gives very strong
support to the idea that inflation may indeed be responsible for both
the CMB anisotropies and the large-scale structure of the universe.
Different models of inflation have different specific predictions for
the fine details associated with the spectrum generated during
inflation. It is these minute differences that will allow cosmologists
to differentiate bewteen alternative models of inflation and discard
those that do not agree with observations. However, most importantly,
perhaps, the pattern of anisotropies predicted by inflation is
completely different from those predicted by alternative models of
structure formation, like cosmic defects: strings, vortices, textures,
etc. These are complicated networks of energy density concentrations
left over from an early universe phase transition, analogous to the
defects formed in the laboratory in certain kinds of liquid crystals
when they go through a phase transition. The cosmological defects have
spectral properties very different from those generated by
inflation. That is why it is so important to launch more sensitive
instruments, and with better angular resolution, to determine the
properties of the CMB temperature and polarization anisotropies. With
the recent observations of these anisotropies by the Microwave
Anisotropy Probe (MAP) satellite, launched by NASA in 2000, we can now
discard topological defects as the source of structure in the universe
at more than ten standard deviations. The full sky coverage of MAP and
its extraordinary angular resolution (10 arcminutes) allows
cosmologists to determine today a handful of cosmological parameters
at the 1\% level, see table 1. We have thus entered the era of
precision cosmology and we can now speak of a truly Standard Model of
Cosmology.

In the next few years, a third generation satellite -- the Planck
Surveyor, due to be launched by the European Space Agency in 2007 --
will measure those temperature anisotropies with 10 times better
angular resolution and 10 times better sensitivity than MAP, and thus
allow cosmologists to determine the parameters of the standard
cosmological model with 1 per mil accuracy. What makes the microwave
background observations particularly powerful is the absence of large
systematic errors that plague other cosmological measurements. As we
have discussed above, the physics of the microwave background is
relatively simple, compared to, say, the physics of supernova
explosions, and computations can be done consistently within
perturbation theory. Thus, most of the systematic errors are
theoretical in nature, due to our ignorance about the primordial
spectrum of metric perturbations from inflation. There is a great
effort at the moment in trying to cover a large region in the
parameter space of models of inflation, to ensure that we have
considered all possible alternatives, like isocurvature or pressure
perturbations, non scale invariant or tilted spectra and non-Gaussian
density perturbations.

In particular, inflation also predicts a spectrum of gravitational
waves. Their amplitude is directly proportional to the total energy
density during inflation, and thus its detection would immediately
tell us about the energy scale (and, therefore, the epoch in the early
universe) at which inflation occurred. If the period of inflation
responsible for the observed CMB anisotropies is associated with the
Grand Unification scale, 12 orders of magnitude above the electroweak
scale, when the strong and electroweak interactions are supposed to
unify, then there is a chance that we might see the effect of
gravitational waves in the future satellite measurements, specially
from the analysis of polarization anisotropies in the microwave
background maps.

Moreover, the stochastic background of gravitational waves generated
during inflation could eventually be observed by ground-based laser
interferometers like LIGO and VIRGO, which will start taking data as
gravitational wave observatories in the next few years. These are
extremely sensitive devices that could distinguish minute spatial
variations, of one part in $10^{23}$ or better, induced when a
gravitational wave from a distant source passes through the Earth and
distorts the space-time metric.  Gravitational waves moving at the
speed of light are a fundamental prediction of general
relativity. Their existence was indirectly confirmed by Russell
A. Hulse and Joseph H. Taylor, through the precise observations of the
decay in the orbital period of the pulsar PSR1913+16, due to the
emission of gravitational radiation. In the near future, observations
of gravitational waves with laser interferometers will open a
completely new window into the universe. It will allow us to observe
with a very different probe (that of the gravitational interaction) a
huge range of phenomena, from the most violent processes in our galaxy
and beyond, like supernova explosions, neutron star collisions,
quasars, gamma ray bursts, etc., to the origin of the universe.
Moreover, NASA and ESA have joined efforts to construct LISA, an
interferometer in space, with satellites millions of kilometers apart,
whose sensitivity is good enough to detect the minutest perturbations
in space-time induced by the stochastic background of gravitational
waves coming from inflation.

In our quest for the parameters of the standard cosmological model,
various groups are searching for distant astrophysical objects that
can serve as standard candles to determine the distance to the object
from their observed apparent luminosity. A candidate that has recently
been exploited with great success is a certain type of supernova
explosions at large redshifts. These are stars at the end of their
life cycle that become unstable and violently explode in a natural
thermonuclear explosion that out-shines their progenitor galaxy. The
intensity of the distant flash varies in time, it takes about three
weeks to reach its maximum brightness and then it declines over a
period of months.  Although the maximum luminosity varies from one
supernova to another, depending on their original mass, their
environment, etc., there is a pattern: brighter explosions last longer
than fainter ones. By studying the light curves of a reasonably large
statistical sample, cosmologists from two competing groups, the
Supernova Cosmology Project and the High-redshift Supernova Project,
are confident that they can use this type of supernova as a standard
candle. Since the light coming from some of these rare explosions has
travelled for a large fraction of the size of the universe, one
expects to be able to infer from their distribution the spatial
curvature and the rate of expansion of the universe. One of the
surprises revealed by these observations is that the universe appears
to be accelerating instead of decelerating, as was expected from the
general attraction of matter; something seems to be acting as a
repulsive force on very large scales. The most natural explanation for
this is the existence of a cosmological constant, a diffuse vacuum
energy that permeates all space and, as explained above, gives the
universe an acceleration that tends to separate gravitationally bound
systems from each other. The origin of such a vacuum energy is one of
the biggests problems of modern physics. Its observed value is 120
orders of magnitude smaller than predicted by quantum mechanics. If
confirmed, it will pose a real challenge to theoretical physics, one
that may affect its most basic foundations. 

\section{The origin of matter in the universe}

Cosmological inflation may be responsible for the metric perturbations
that later gave rise to the large scale structures we see in the
universe, but where did all the matter in the universe come from? Why
isn't all the energy in photons, which would have inevitably redshifted
away in a cold universe devoid of life? How did we end up being matter
dominated?  Everything we see in the universe, from planets and stars,
to galaxies and clusters of galaxies, is made out of matter, so where
did the antimatter in the universe go? Is this the result of an
accident, a happy chance occurrence during the evolution of the
universe, or is it an inevitable consequence of some asymmetry in the
laws of nature?  Theorists believe that the excess of matter over
antimatter comes from fundamental differences in their interactions soon
after the end of inflation.

Inflation is an extremely efficient mechanism in diluting any parti\-cle
species or fluctuations. At the end of inflation, the universe is empty
and extremely cold, dominated by the homogeneous coherent mode of the
inflaton. Its potential energy density is converted into particles, as
the inflaton field oscillates coherently around the minimum of its
potential, see Fig~\ref{fig2}. These particles are initially very far from
equilibrium, but they strongly interact among themselves and soon reach
thermal equilibrium at a very large temperature. From there on, the
universe expanded isoentropically, cooling down as it expanded, in the
way described by the standard hot Big Bang model. Thus the origin of the
Big Bang itself, and the matter and energy we observe in the universe
today, can be traced back to the epoch in which the inflaton energy
density decayed into particles. Such a process is called reheating of
the universe. 

Recent developments in the theory of reheating suggest that the decay of
the inflaton energy could be explosive due to the coherent oscillations
of the inflaton, which induce its stimulated decay. The result is a
resonant production of particles in just a few inflaton oscillations, an
effect very similar to the stimulated emission of a laser beam of
photons. The number of particles produced this way is exponentially
large, which may explain the extraordinarily large entropy, of order
$10^{89}$ particles, in our observable patch of the universe today.
However, the inflaton is supposed to be a neutral scalar field, and thus
its interactions cannot differentiate between particles and
antiparticles. How did we end up with more matter than antimatter? The
study of this cosmological asymmetry goes by the name of baryogenesis
since baryons (mainly protons and neutrons) are the fundamental
constituents of matter in planets, stars and galaxies in the universe
today. So, what are the conditions for baryogenesis?

Everything we know about the properties of elementary particles is
included in the standard model of particle physics. It describes more
than 100 observed particles and their interactions in terms of a few
fundamental constituents: six quarks and six leptons, and their
antiparticles. The standard model describes three types of interactions:
the electromagnetic force, the strong and the weak nuclear forces. These
forces are transmitted by the corresponding particles: the photon, the
gluon and the W and Z bosons. The theory also requires a scalar
particle, the Higgs particle, responsible for the masses of quarks and
leptons and the breaking of the electroweak symmetry at an energy scale
1000 times the mass of the proton. The Higgs is believed to lie behind
most of the mysteries of the standard model, including possibly also
the asymmetry between matter and antimatter.

In 1967, the Russian physicist Andrei Sakharov pointed out the three
necessary conditions for the baryon asymmetry of the universe to
develop. First, we need interactions that do not conserve baryon
number B, otherwise no asymmetry could be produced in the first
place. Second, C and CP symmetry must be violated, in order to
differentiate between matter and antimatter, otherwise B
non-conserving interactions would produce baryons and antibaryons at
the same rate, thus maintaining zero net baryon number.  Third, these
processes should occur out of thermal equilibrium, otherwise particles
and antiparticles would be produced at the same rate.  The standard
model is baryon symmetric at the classical level, but violates B at
the quantum level, through the chiral anomaly. Electroweak
interactions violate C and CP, but the magnitude of the latter is
clearly insufficient to account for the observed baryon
asymmetry. This failure suggests that there must be other sources of
CP violation in nature, and thus the standard model of particle
physics is probably incomplete.

One of the most popular extensions of the standard model includes a
new symmetry called supersymmetry, which relates bosons (particles
that mediate interactions) with fermions (the constituents of
matter). Those extensions generically predict other sources of CP
violation coming from new interactions at scales above 1000 times the
mass of the proton. Such scales will soon be explored by particle
colliders like the Large Hadron Collider (LHC) at CERN (the European
Centre for Particle Physics) and by the Tevatron at Fermilab.  The
mechanism for baryon production in the early universe in these models
relies on the strength of the electroweak phase transition, as the
universe cooled and the symmetry was broken. Only for strongly
first-order phase transitions is the universe sufficiently far from
equilibrium to produce enough baryon asymmetry. Unfortunately, the
phase transition in these models is typically too weak to account for
the observed asymmetry, so some other mechanism is needed.

If reheating after inflation occurred in an explosive way, via the
resonant production of particles from the inflaton decay, as recent
developments suggest, then the universe has actually gone through a very
non-linear, non-perturbative and very far from equilibrium stage, before
thermalizing via particle interactions. Electroweak baryogenesis could
then take place during that epoch, soon after the end of low energy
inflation at the electroweak scale. Such models can be constructed but
require a specially flat direction (a very small mass for the inflaton)
during inflation, in order to satisfy the constraints from the amplitude
of temperature anisotropies seen by COBE. Such flat directions are
generic in supersymmetric extensions of the standard model. After
inflation, the inflaton acquires a large mass from its interaction with
the Higgs field.

The crucial ingredient of departure from equilibrium, necessary for
the excess production of baryons over antibaryons, is strongly present
in this new scenario of baryogenesis, as the universe develops from a
zero-temperature and zero-entropy state, at the end of inflation, to a
thermal state with exponentially large numbers of particles, the
origin of the standard hot Big Bang. If, during this stage,
fundamental or effective interactions that are B, C and CP violating
were fast enough compared to the rate of expansion, the universe could
have ended with the observed baryon asymmetry of one part in
$10^{10}$, or one baryon per $10^9$ photons today, as deduced from
observations of the light element abundances. Recent calculations
suggest than indeed, the required asymmetry could be produced as long
as some new physics, just above the electroweak symmetry breaking
scale, induces a new effective CP violating interaction.

These new phenomena necessarily involve an interaction between the
Higgs particle, responsible for the electroweak symmetry breaking, and
the inflaton field, responsible for the period of cosmological
inflation.  Therefore, for this scenario to work, it is expected that
both the Higgs and the inflaton particles be discovered at the future
particle physics colliders like the LHC and the Next Linear Collider
(NLC). Furthermore, this new physics would necessarily involve new
interactions in the quark sector, for example inducing CP violations
in the B meson (a bound state composed of a bottom quark and an
antidown quark) system.  Such violations are the main research
objective of the B factory at SLAC in California and at KEK, the High
Energy Accelerator Research Organization in Tsukuba, Japan. These
experiments have already been collecting data for a couple years, and
for the moment are in perfect agreement with the Standard Model of
particle physics. However, perhaps in the near future they may detect
a deviation which could give us a clue to the origin of CP, and thus
to the matter--antimatter asymmetry of the Universe and, possibly, to
baryogenesis from reheating after inflation.

\section{Conclusions}

We have entered a new era in cosmology, were a host of high-precision
measurements are already posing challenges to our understanding of the
universe: the density of ordinary matter and the total amount of energy
in the universe; the microwave background anisotropies on a fine-scale
resolution; primordial deuterium abundance from quasar absorption lines;
the acceleration parameter of the universe from high-redshift supernovae
observations; the rate of expansion from gravitational lensing; large
scale structure measurements of the distribution of galaxies and their
evolution; and many more, which already put constraints on the parameter
space of cosmological models. However, these are only the
forerunners of the precision era in cosmology that will dominate this
millennium, and will make cosmology a science in its own right.

It is important to bear in mind that all physical theories are
approximations of reality that can fail if pushed too far. Physical
science advances by incorporating earlier theories that are
experimentally supported into larger, more encompassing frameworks. The
standard Big Bang theory is supported by a wealth of evidence, nobody
really doubts its validity anymore. However, in the last decade it has
been incorporated into the larger picture of cosmological inflation,
which has become the new standard cosmological model. All cosmological
issues are now formulated in the context of the inflationary paradigm.
It is the best explanation we have at the moment for the increasing
set of cosmological observations.

In the next few years we will have an even larger set of high-quality
observations that will test inflation and the cold dark matter
paradigm of structure formation, and determine most of the 20 or more
parameters of the standard cosmological model to a few per mil
accuracy, see Table~1. It may seem that with such a large number of
parameters one can fit almost anything. However, that is not the case
when there is enough quantity and quality of data. An illustrative
example is the standard model of particle physics, with around 21
parameters and a host of precise measurements from particle
accelerators all over the world. This model is, nowadays, rigurously
tested, and its parameters measured to a precision of better than 1\%
in most cases. It is clear that high-precision measurements will make
the standard model of cosmology as robust as that of particle
physics. This is definitely a very healthy field, but there is still a
lot to do. With the advent of better and larger precision experiments,
cosmology is becoming a mature science, where speculation has given
way to phenomenology.

However, there are still many unanswered fundamental questions in this
emerging picture of cosmology. For instance, we still do not know the
nature of the inflaton field, is it some new fundamental scalar field
in the electroweak symmetry breaking sector, or is it just some
effective description of a more fundamental high energy interaction?
Hopefully, in the near future, experiments in particle physics might
give us a clue to its nature. Inflation had its original inspiration
in the Higgs field, the scalar field supposed to be responsible for
the masses of elementary particles (quarks and leptons) and the
breaking of the electroweak symmetry. Such a field has not been found
yet, and its discovery at the future particle colliders may help
understand one of the truly fundamental problems in physics, the
origin of masses. If the experiments discover something completely new
and unexpected, it would automatically affect inflation at a
fundamental level.

One of the most difficult challenges that the new cosmology will have
to face is understanding the origin and nature of the cosmological
constant. Ever since Einstein introduced it as a way to counteract
gravitational attraction, it has haunted cosmologists and particle
physicists. We still do not have a mechanism to explain its
extraordinarily small value, 120 orders of magnitude below what is
predicted by quantum physics. For several decades there has been the
reasonable speculation that this fundamental problem may be related to
the quantization of gravity.  General relativity is a classical theory
of space-time, and it has proved particularly difficult to construct a
consistent quantum theory of gravity, since it involves fundamental
issues like causality and the nature of space-time itself.

The value of the cosmological constant predicted by quantum physics is
related to our lack of understanding of gravity at the microscopic
level. However, its effect is dominant at the very largest scales of
clusters or superclusters of galaxies, on truly macroscopic scales. We
can speculate that perhaps general relativity is not the correct
description of gravity on the very largest scales. In fact, it is only
in the last few billion years that the observable universe has become
large enough that these global effects could be noticeable. In its
infancy, the universe was much smaller than it is now, and,
presumably, general relativity gave a correct description of its
evolution, as confirmed by the successes of the standard Big Bang
theory. As it expanded, larger and larger regions were encompassed,
and, therefore, deviations from general relativity would slowly become
important. It may well be that the recent determination of a
cosmological constant from observations of supernovae at high
redshifts is hinting at a fundamental misunderstanding of gravity on
the very large scales. If this were indeed the case, we should expect
that the new generation of precise cosmological observations will not
only affect our cosmological model of the universe but also a more
fundamental description of nature.

\end{document}